\documentclass[10pt,apjl,tighten,iop]{emulateapj}

\usepackage{amsmath}   % for \being{align}...
\usepackage{graphicx}
\usepackage{tabulary}
\usepackage{epsfig}
\usepackage{amssymb}
\usepackage{multirow}
\usepackage{appendix}
\usepackage{natbib}
\usepackage{lineno}
\usepackage{wasysym}
\usepackage[pdftitle={How to determine an exomoon's sense of orbital motion}, colorlinks = true, breaklinks = true,
  citecolor = blue, linkcolor = blue, urlcolor = blue, pdfauthor =
  {Ren\'{e} Heller}]{hyperref}
\newcommand{\beq}[1]{\begin{equation}\label{#1}}
\newcommand{\eeq}{\end{equation}}

\shorttitle{How to determine an exomoon's sense of orbital motion}

\shortauthors{Ren\'e Heller {{\&}} Simon Albrecht}

\begin{document}

%FRONT MATTER
\title{How to determine an exomoon's sense of orbital motion}
\author{Ren\'e Heller\altaffilmark{1,2}}
\affil{Origins Institute, McMaster University, Hamilton, ON L8S 4M1, Canada; \href{mailto:rheller@physics.mcmaster.ca}{rheller@physics.mcmaster.ca}}

\and

\author{Simon Albrecht}
\affil{Stellar Astrophysics Centre, Department of Physics and Astronomy, Aarhus University, Ny Munkegade 120, DK-8000 Aarhus C, Denmark; \href{mailto:albrecht@phys.au.dk}{albrecht@phys.au.dk}}

\altaffiltext{1}{Department of Physics and Astronomy, McMaster University}
\altaffiltext{2}{Postdoctoral fellow of the Canadian Astrobiology Training Program}

%ABSTRACT
\begin{abstract}
We present two methods to determine an exomoon's sense of orbital motion (SOM), one with respect to the planet's circumstellar orbit and one with respect to the planetary rotation. Our simulations show that the required measurements will be possible with the European Extremely Large Telescope (E-ELT). The first method relies on mutual planet-moon events during stellar transits. Eclipses with the moon passing behind (in front of) the planet will be late (early) with regard to the moon's mean orbital period due to the finite speed of light. This ``transit timing dichotomy'' (TTD) determines an exomoon's SOM with respect to the circumstellar motion. For the ten largest moons in the solar system, TTDs range between 2 and 12\,s. The E-ELT will enable such measurements for Earth-sized moons around nearby stars. The second method measures distortions in the IR spectrum of the rotating giant planet when it is transited by its moon. This Rossiter-McLaughlin effect (RME) in the planetary spectrum reveals the angle between the planetary equator and the moon's circumplanetary orbital plane, and therefore unveils the moon's SOM with respect to the planet's rotation. A reasonably large moon transiting a directly imaged planet like $\beta$\,Pic\,b causes an RME amplitude of almost $100\,\mathrm{m\,s}^{-1}$, about twice the stellar RME amplitude of the transiting exoplanet HD209458\,b. Both new methods can be used to probe the origin of exomoons, that is, whether they are regular or irregular in nature.
\end{abstract}

%KEY WORDS

\section{CONTEXT}
\label{sec:context}

Although thousands of extrasolar planets and candidates have been found, some as small as the Earth's Moon \citep{2013Natur.494..452B}, no extrasolar moon has been detected. The first dedicated hunts for exomoons have now been initiated \citep{2007AaA...476.1347P,2012ApJ...750..115K,2013AaA...553A..17S}, and it has been shown that the Kepler or PLATO space telescopes may find large exomoons in the stellar light curves \citep{2009MNRAS.400..398K,2014ApJ...787...14H}, if such worlds exist.

The detection of exomoons would be precious from a planet formation perspective, as giant planet satellites carry information about the thermal and compositional properties in the early circumplanetary accretion disks \citep{2006Natur.441..834C,HellerPudritz14}. Moons can also constrain the system's collision history \citep[see the Earth-Moon binary,][]{1975Icar...24..504H} and bombardment record \citep[see the misaligned Uranian system,][]{2012Icar..219..737M}, they can trace planet-planet encounters \citep[see Triton's capture around Neptune,][]{2006Natur.441..192A}, and even the migration history of close-in giant planets \citep{2010ApJ...719L.145N}. Under suitable conditions, an exomoon observation could reveal the absolute masses and radii in a star-planet-moon system \citep{2010MNRAS.409L.119K}. What is more, moons may outnumber rocky planets in the stellar habitable zones \citep{2013arXiv1311.0292H} and therefore could be the most abundant species of habitable worlds \citep{1997Natur.385..234W,2014arXiv1408.6164H}.

A moon's sense of orbital motion (SOM) is crucial to determine its origin and orbital history. About a dozen techniques have been proposed to find an extrasolar moon \citep{2014ApJ...787...14H}, but none of them can determine an exomoon's SOM with current technical equipment \citep{2014ApJ...791L..26L}. We here identify two means to determine an exomoon's SOM relative to the circumstellar orbit and with respect to the planet's direction of rotation. In our simulations, we use the European Extremely Large Telescope\footnote{Construction of the E-ELT near the Paranal Observatory in Chile began in June 2014, with first light anticipated in 2024.} (E-ELT) as an example for one of several ELTs now being built.

\section{METHODS}
\label{sec:methods}

%**********************************************
%Fig. 1
\begin{figure}[t]
  \centering
  \vspace{.0cm}
  \scalebox{0.22}{\includegraphics{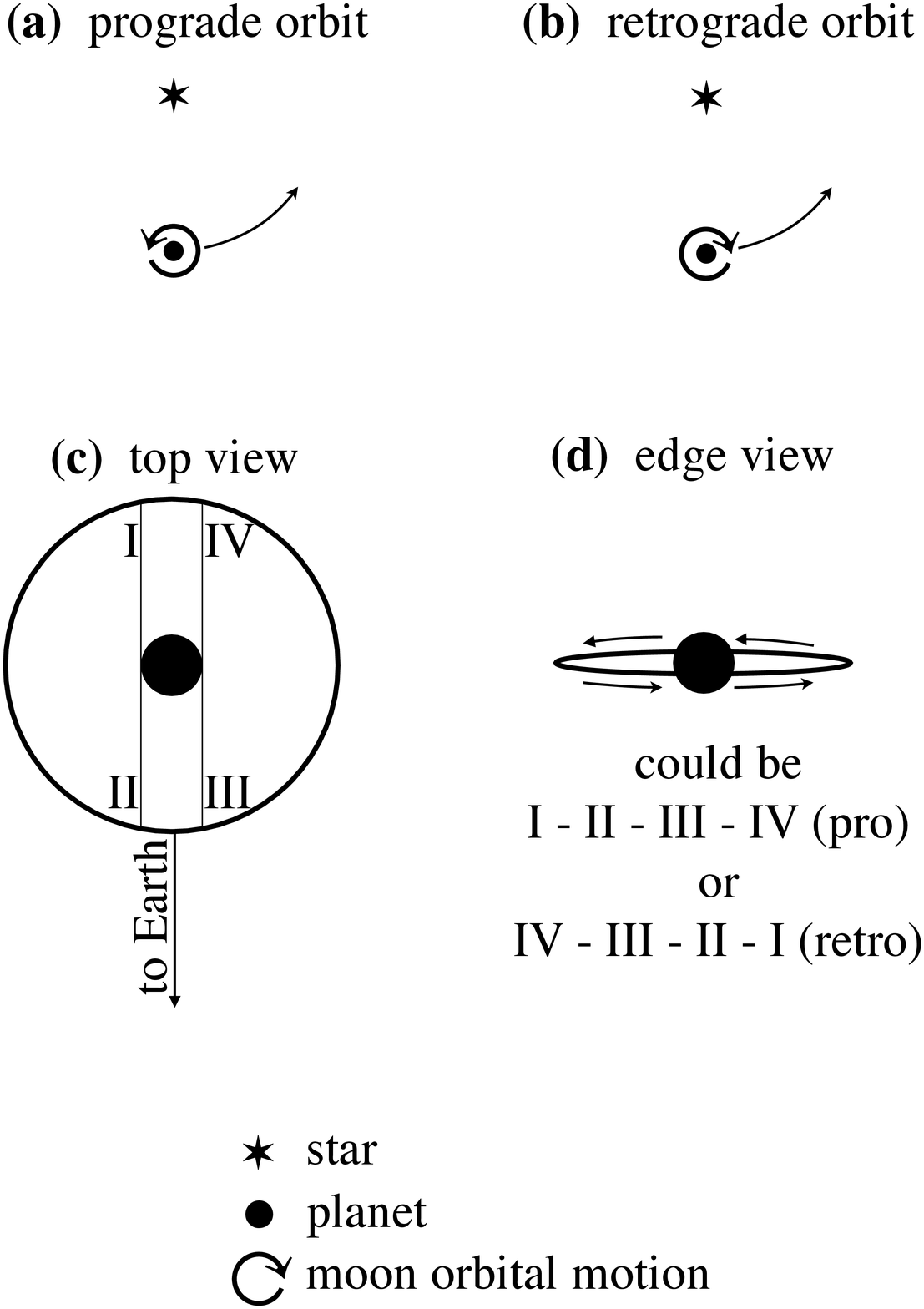}}\\
  \scalebox{0.15}{\includegraphics{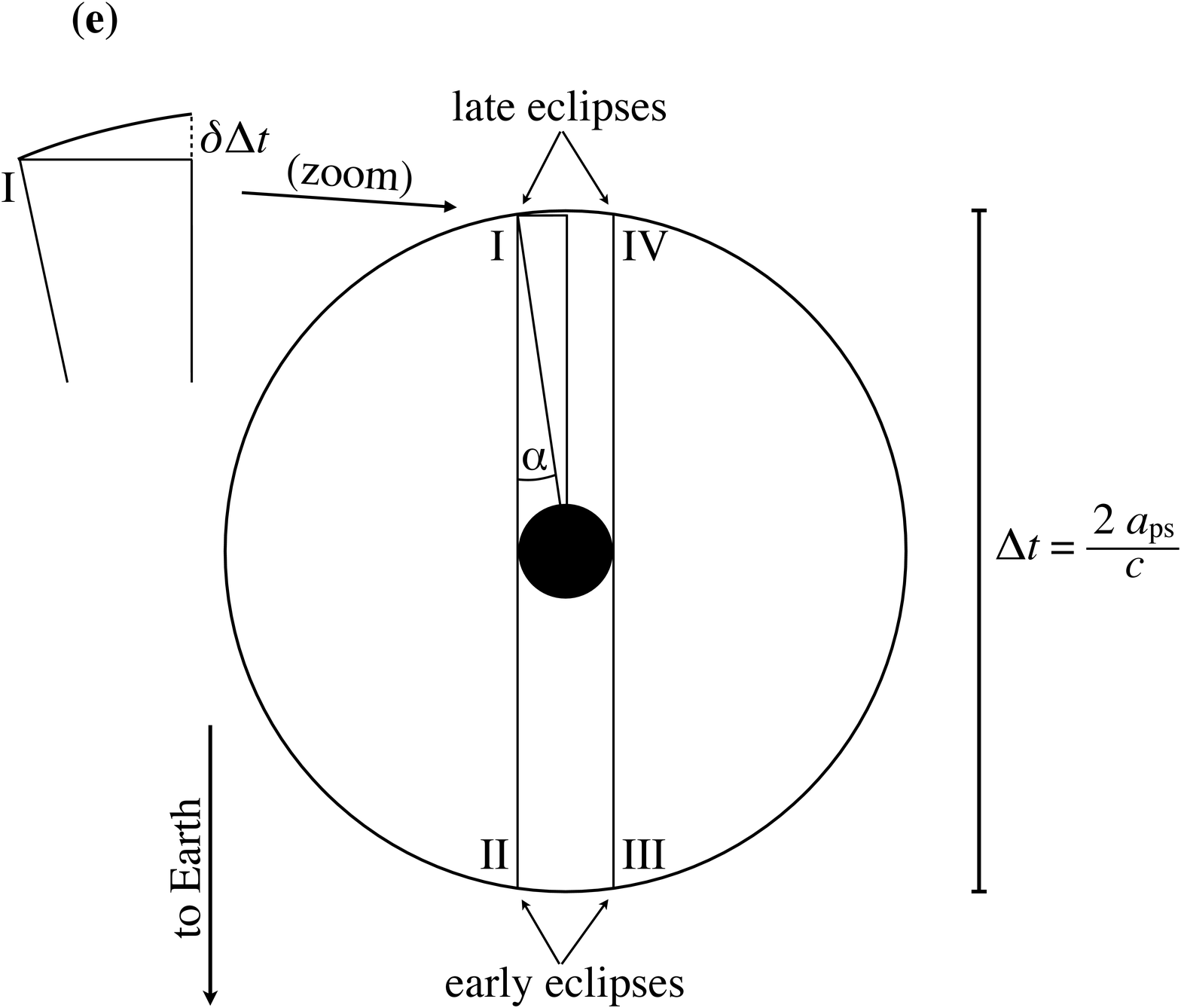}}
  \caption{Orbital geometry of a star-planet-moon system in circular orbits. {\bf (a)} Top view of the system's orbital motion. The moon's circumplanetary orbit is prograde with respect to the circumstellar orbit. {\bf (b)} Similar to {\bf (a)}, but now the moon is retrograde. {\bf (c)} Top view of the moon's orbit around the planet. Roman numbers I to IV denote the ingress and egress of mutual planet-moon eclipses. {\bf (d)} Edge view (as seen from Earth) of a circumplanetary moon orbit. {\bf (e)} Transit timing dichotomy of mutual planet-moon events. Due to the finite speed of light, an Earth-bound observer witnesses events I and IV with a positive time delay ${\Delta}t$ compared to events II and III, respectively.}
     \vspace{.15cm}
  \label{fig:geometry}
\end{figure}
%**********************************************

\subsection{An Exomoon's Transit Timing Dichotomy (TTD)}

For our first new method to work, the moon needs to be large enough (and the star's photometric variability sufficiently low) to cause a direct transit signature in the stellar light curve \citep{1999AaAS..134..553S}. Depending on the moon's orbital semi-major axis around the planet ($a_\mathrm{ps}$) and on the orbital alignment, some stellar transits of the planet-moon pair will then show mutual planet-moon eclipses. These events have been simulated \citep{2007AaA...464.1133C,2009PASJ...61L..29S,2011MNRAS.416..689K,2012MNRAS.420.1630P}, and a planet-planet eclipse \citep{2012ApJ...759L..36H} as well as mutual events in a stellar triple system \citep{2011Sci...331..562C} have already been found in the Kepler data.

Figure~\ref{fig:geometry} illustrates the difficulty in determining a moon's SOM. Panels (a) and (b) visualize the two possible scenarios of a prograde and a retrograde SOM with respect to the circumstellar motion. Panel (c) presents the four possible ingress and egress locations (arbitrarily labelled I, II, III, and IV) for a mutual planet-moon event during a stellar transit. Panel (d) shows the projection of the three-dimensional moon orbit on the two-dimensional celestial plane. If the moon transit is directly visible in the stellar light curve, then events I and II can be distinguished from events III and IV, simply by determining whether the moon enters the stellar disk first and then performs a mutual event with the planet (III and IV) or the planet enters the stellar disk first before a mutual event (I and II). However, this inspection cannot discern event I from II or event III from IV. Therefore, prograde and retrograde orbits cannot be distinguished from each other.

Figure~\ref{fig:geometry}~(e) illustrates how the I/II and III/IV ambiguities can be solved. Due to the finite speed of light, there will be a time delay between events I and II, and between events III and IV. It will show up as a transit timing dichotomy (TTD) between mutual events where the moon moves in front of the planetary disk (early mutual events) or behind it (late mutual events). Events I and IV will be late by ${\Delta}t=2a_\mathrm{ps}/c$ compared to events II and III, respectively. Imagine that two mutual events, either I and [II or III] or IV and [III or II]), are observed during two different stellar transits and that the moon has completed $n$ circumplanetary orbits between the two mutual events. Then it is possible to determine the sequence of late and early eclipses, and therefore the SOM with respect to the circumstellar movement, if (1) the orbital period of the planetary satellite ($P_\mathrm{ps}$) can be determined independently with an accuracy ${\delta}P_\mathrm{ps}<P_\mathrm{ps}/n$, and if (2) the event mid-times can be measured with a precision $<{\Delta}t$. As a byproduct, measurements of $\Delta{t}$ yield an estimate for $a_\mathrm{ps}=\Delta{t}\times{c}$.

Concerning (1), the planet's transit timing variation (TTV) and transit duration variation (TDV) due to the moon combined may constrain the satellite mass ($M_\mathrm{s}$) and $P_\mathrm{ps}$ \citep{2011tepm.book.....K}. As an  example, if two mutual events of a Jupiter-Ganymede system at 0.5\,AU around a Sun-like star were observed after 10 stellar transits (or 3.5\,yr), the moon ($P_\mathrm{ps}\approx0.02$\,yr) would have completed $n\approx175$ circumplanetary orbits. Hence, ${\delta}P_\mathrm{ps}\lesssim1\,\mathrm{hr}$ would be required. A combination of TTV and TDV measurements might be able to deliver such an accuracy \citep[Section~6.5.1 in][]{2011tepm.book.....K}.

For our estimates of $\Delta{t}$, we can safely approximate that a moon enters a mutual event at a radial distance $a_\mathrm{ps}$ to the planet, because

\begin{align}\label{eq:TTD} \nonumber
\delta\Delta{t} &= \frac{a_\mathrm{ps}}{c} {\Big (} 1-\cos(\alpha) {\Big )} \nonumber \\
                       &= \frac{a_\mathrm{ps}}{c} {\Bigg (}1-\cos{\Big \{}  \arcsin{\Big (}\frac{R_\mathrm{p}}{a_\mathrm{ps}}{\Big )} {\Big \}} {\Bigg )} \nonumber \\
                       & \ll \Delta{t} \, \, ,
\end{align}

\noindent
with $c$ being the speed of light, $R_\mathrm{p}$ the planetary radius, and $\alpha$ defined by $\sin(\alpha)=R_{\rm p}/a_{\rm ps}$ as shown in Figure~\ref{fig:geometry}(e). For a moon at $15\,R_\mathrm{p}$ from its planet, such as Ganymede around Jupiter, $\alpha\approx3.8^\circ$ and $\delta\Delta{t}\approx0.005$\,s, which is completely negligible. Orbital eccentricities could also cause light travel times different from the one shown in Figure~\ref{fig:geometry}. But even for eccentricities comparable to Titan's value around Saturn, with 0.0288 the largest among the major moons in the solar system, TTDs would be affected by $<5\,\%$, or fractions of a second. Significantly larger eccentricities are unlikely, as they will be tidally eroded within a million years \citep{2011ApJ...736L..14P,2013AsBio..13...18H}.

\subsection{The Rossiter-McLaughlin effect (RME) in the Planetary Emission Spectrum}

%**********************************************
%Fig. 2
\begin{figure}[t]
  \centering
  \vspace{.0cm}
  \scalebox{0.608}{\includegraphics{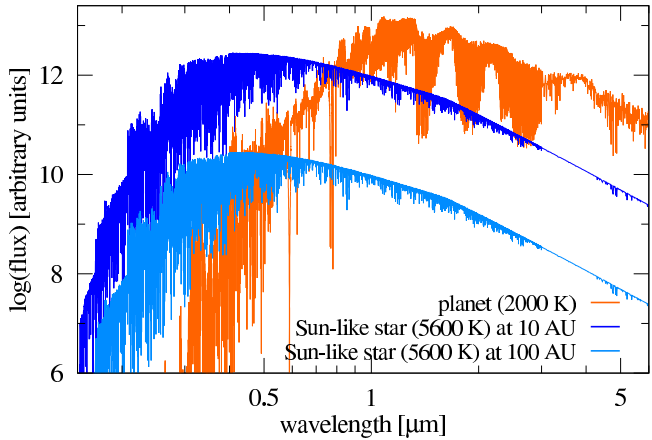}}
  \caption{Theoretical emission spectra of a hot, Jupiter-sized planet (orange, upper) and a Sun-like star as reflected by the planet at 10\,AU (blue, middle) and 100\,AU (light blue, lower). For the planet, a Jupiter-like bond albedo of 0.3 is assumed.}
     \vspace{.2cm}
  \label{fig:spectra}
\end{figure}
%**********************************************

In the solar system, all planets except Venus \citep{1969Icar...11..356G} and Uranus rotate in the same direction as they orbit the Sun. One would expect that the orbital motion of a moon is aligned with the rotation of the planet. However, collisions, gravitational perturbations, capture scenarios etc. can substantially alter a satellite's orbital plane \citep{2014arXiv1408.6164H}. Hence, knowledge about the spin-orbit misalignment, or obliquity, in a planet-exomoon system would be helpful in inferring its formation and evolution.

One such method to constrain obliquities is the Rossiter-McLaughlin effect (RME), a distortion in the rotationally broadened absorption lines caused by the partial occultation of the rotating sphere (typically a star) by a transiting body (usually another star or a planet). This distortion can either be measured directly \citep{2007AaA...474..565A,2010MNRAS.403..151C} or be picked up as an RV shift during transit. The shape of this anomalous RV curve reveals the projection of the angle between the orbital normal of the occulting body (in our case the moon) and the rotation axis of the occulted body (here the planet). Originally observed in stellar binaries \citep{1924ApJ....60...15R,1924ApJ....60...22M}, this technique experienced a renaissance in the age of extrasolar planets, with the first measurement taken by \citet{2000AaA...359L..13Q} for the transiting hot Jupiter HD209458\,b. Numerous measurements\footnote{So far, 79 exoplanet have had their spin-orbit alignment measured via the RME, see R.~Heller's Holt-Rossiter-McLaughlin Encyclopaedia (\href{http://www.physics.mcmaster.ca/~rheller}{www.physics.mcmaster.ca/$\sim$rheller}).} revealed planets in aligned, misaligned, and even retrograde orbits -- strongly contrasting the architecture of the solar system \citep[e.g.][]{2012ApJ...757...18A}. The \textit{stellar} RME of exomoons has been studied before \citep{2010MNRAS.406.2038S,2012ApJ...758..111Z}, but here we refer to the RME in the \textit{planetary} infrared spectrum caused by the moon transiting a hot, young giant planet.

%**********************************************
%Fig. 3
\begin{figure}[t]
  \centering
  \vspace{.0cm}
  \scalebox{0.608}{\includegraphics{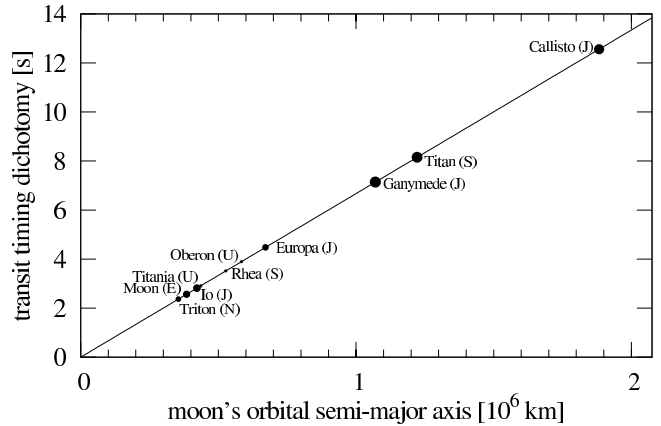}}
  \caption{Transit timing dichotomies of the ten largest moons in the solar system. Moon radii are symbolized by circle sizes. The host planets Jupiter, Saturn, Neptune, Uranus, and Earth are indicated with their initials. Note that the largest moons, causing the deepest solar transits, induce the highest TTDs.}
     \vspace{.1cm}
  \label{fig:TTD}
\end{figure}
%**********************************************

For this method to be effective, a giant planet's light needs to be measured directly. Starting with the planetary system around HR\,8799 \citep{2008Sci...322.1348M} and the planet candidate Fomalhaut\,b \citep{2008Sci...322.1345K}, 18 giant exoplanets have now been directly imaged, most of which are hot ($>1000$\,K), and young ($<100$\,Myr). Upcoming instruments like SPHERE \citep{2006Msngr.125...29B} and GPI \citep{2011PASP..123..692M} promise a rapid increase of this number. Stellar and planetary spectra can also be separated in velocity space without the need of spatial separation \citep{2012Natur.486..502B,2013AaA...554A..82D,2013MNRAS.436L..35B}, but for $\beta$\,Pic-like systems, instruments like CRIRES can also separate stellar and planetary spectra spatially. \citet{2014Natur.509...63S} determined the rotation period of $\beta$\,Pic\,b to be $8.1\pm1.0$ hours. The high rotation velocity ($\approx25\,\mathrm{km\,s}^{-1}$) favors a large RME amplitude, but makes RV measurements more difficult.

Contamination of the planetary spectrum by the star via direct stellar light on the detector and stellar reflections from the planet might pose a challenge. Consequently, observations need to be carried out in the near-IR, where the planet is relatively bright and presents a rich forest of spectral absorption lines. Figure~\ref{fig:spectra} shows that contamination of the planetary spectrum\footnote{Models were provided by T.-O. Husser (priv.~comm.), based on the spectral library by \citet{2013AaA...553A...6H} available at \href{http://phoenix.astro.physik.uni-goettingen.de/}{http://phoenix.astro.physik.uni-goettingen.de}.} by reflected star light becomes negligible beyond several 10\,AU, with no need for additional cleaning.

\section{Results and Predictions}
\label{sec:predictions}

\subsection{Transit Timing Dichotomies}

%**********************************************
%Fig. 4
\begin{figure}[t]
  \centering
  \vspace{.0cm}
  \scalebox{0.6}{\includegraphics{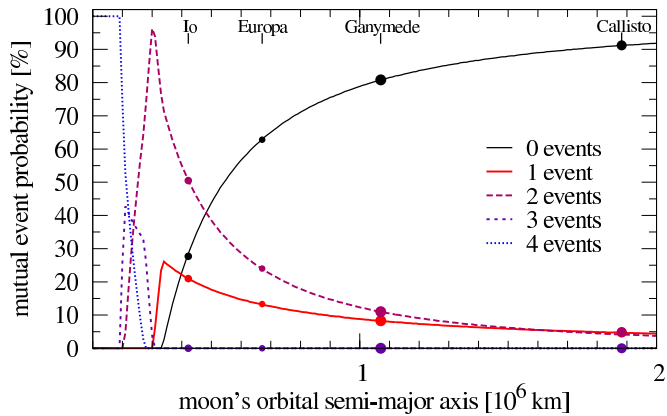}}
  \caption{Probability of mutual planet-moon events as a function of the moon's planetary distance around a Jupiter-like planet, which is assumed to orbit a Sun-like star at 1\,AU. The five curves indicate the frequency of 0, 1, 2, 3, or 4 mutual events during stellar stellar transits as measured in our transit simulations. The orbits of Io, Europa, Ganymede, and Callisto are indicated with symbols along each curve.}
     \vspace{.2cm}
  \label{fig:prob}
\end{figure}
%**********************************************

We computed the TTDs of the ten largest moons in the solar system, yielding values between about 2 and 12\,s (Figure~\ref{fig:TTD}). Most intriguingly, the largest moons (Ganymede, Titan, and Callisto), which have the deepest solar transit signatures, also have the largest TTDs. This is owed to the location of the water ice line in the accretion disks around Jovian planets, which causes the most massive icy satellites to form beyond about $15\,R_\mathrm{Jup}$ \citep{HellerPudritz14}. Figure~\ref{fig:TTD} indicates that timing precisions of 1 - 6\,s need to be achieved on transit events with depths of only about $10^{-4}$, corresponding to the transit depth of an Earth-sized moon transiting a Sun-like star.

Precisions of 6\,s in exoplanet transit mid-times have been achieved from the ground using the Baade 6.5\,m telescope at Las Campanas Observatory in Chile \citep{2009AJ....137.3826W}. On the one hand, the planet in these observations (WASP-4b) was comparatively large to its host star with a transit depth of about 2.4\,\%. On the other hand, the star was not particularly bright, with an apparent visual magnitude $m_\mathrm{V}\approx12.6$. The photon collecting power of the E-ELT will be $(39.3\,\mathrm{m}/6.5\,\mathrm{m})^2\approx37$ times that of the Baade telescope. And with improved data reduction methods, timing precisions of the order of seconds should be obtainable for transit depths of $10^{-4}$ with the E-ELT for very nearby transiting systems.

We calculate the probabilities of one to four mutual events ($\mathcal{P}_i$, $i\in\{1,2,3,4\}$) during a stellar transit of a coplanar planet-moon system. The distance of the moon travelled on its circumplanetary orbit during the stellar transit $s=2R_{\star}P_\mathrm{\star b}a_\mathrm{ps}/(P_\mathrm{ps}a_\mathrm{\star b})$, with $a_\mathrm{\star b}$ denoting the circumstellar orbital semi-major axis of the planet-moon barycenter and $P_\mathrm{\star b}$ being its circumstellar orbital period. We analyze a Jupiter-like planet at 1\,AU from a Sun-like star and simulate $10^6$ transits for a range of possible moon semi-major axes, respectively, where the moon's initial orbital position during the stellar transit is randomized. As the stellar transit occurs, we follow the moon's circumplanetary orbit and measure the number of mutual events (of type I, II, III, or IV) during the transit, which can be 0, 1, 2, 3, or even 4.

For siblings of Io, Europa, Ganymede, and Callisto we find $\mathcal{P}_1=21$, 13, 8, and 5\,\% as well as $\mathcal{P}_2=50$, 24, 11, and 5\,\%, respectively (Figure~\ref{fig:prob}). Notably, the probabilities for two mutual events (purple long-dashed line) are higher than the likelihoods for only one (red solid line) if $a_{\rm ps}~{\lesssim}~1.6\times10^6$\,km. For moons inside about half-way between Io and Europa, the probability of having no event (black solid line) is $<50\,\%$, so it is more likely to have at least one mutual event during any transit than having none. Moons inside 200,000\,km (half the semi-major axis of Io) can even have three or more events, allowing the TTD method to work with only one stellar transit. Deviations from well-aligned orbits due to high transit impact parameters or tilted moon orbits would naturally reduce the shown probabilities. Nevertheless, mutual events could obviously be common during transits.

A single stellar transit with $\geq3$ mutual events contains TTD information on its own. A two-event stellar transit only delivers TTD information if the events are either a combination of I and II or of III and IV. Moons in wide orbits cannot proceed from one conjunction to the other during one stellar transit. Hence, the contribution of TTD-containing events along $\mathcal{P}_2$ is zero beyond about $10^6$\,km. In close orbits, the fraction of two-event cases with TTD information is $P_{\rm 2,TTD}{\approx}R_{\rm p}(a_{\rm ps}{\pi})^{-1}$. For Io, as an example, $P_{\rm 2,TTD} = (6.1\times\pi)^{-1}\approx5\,\%$. So even for very close-in moons, two-event cases with TTD information are rare.

\subsection{The Planetary Rossiter-McLaughlin Effect due to an Exomoon}

%**********************************************
%Fig. 5
\begin{figure}[t]
  \centering
  \vspace{.0cm}
  \scalebox{0.508}{\includegraphics{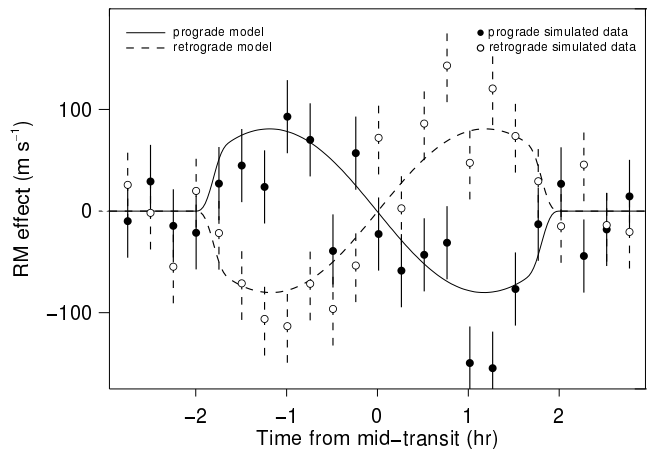}}
  \caption{Simulated Rossiter-McLaughlin effect of a giant moon ($0.7\,R_\oplus$) transiting a hot, Jupiter-sized planet similar to $\beta$\,Pic\,b. Solid and dashed lines correspond to a prograde and a retrograde coplanar orbit, respectively. Full and open circles indicate simulated E-ELT observations.}
     \vspace{-.0cm}
  \label{fig:RME}
\end{figure}
%**********************************************

We simulated the RME in the near-IR spectrum of a planet similar to $\beta$\,Pic\,b assuming a planetary rotation speed of $25\,\mathrm{km\,s}^{-1}$ \citep{2014Natur.509...63S} and a Jovian planetary radius ($R_\mathrm{Jup}$). The moon was placed in a Ganymede-like orbit ($a_{\rm ps}=15\,R_\mathrm{Jup}$) and assumed to belong to the population of giant moons that form at the water ice lines around super-Jovian planets, with a radius of up to about $0.7$ Earth radii ($R_\oplus$) \citep{HellerPudritz14}. Using the code of \citet{2007AaA...474..565A,2013ApJ...771...11A}, we simulated absorption lines of the rotating planet as distorted during the moon's transit with a cadence of 15\,min. Focusing on the same spectral window ($2.304-2.332\,\mu{\rm m}$) as \citet{2014Natur.509...63S}, we then convolved the planetary spectrum (Figure~\ref{fig:spectra}) with these distorted absorption lines. Employing the CRIRES Exposure Time Calculator and incorporating the increase of $(39.3\,\mathrm{m}/8.2\,\mathrm{m})^2\approx23$ in collecting area for the E-ELT, we obtain a S/N of 75 per pixel for a 15\,min exposure -- the same values as obtained by \citet{2014Natur.509...63S} for a similar calculation. The resulting pseudo-observed spectra are finally cross-correlated with the template spectrum, and a Gaussian is fitted to the cross-correlation functions to obtain RVs.

Pro- and retrograde coplanar orbits are clearly distinguishable in the resulting RME curve (Figure~\ref{fig:RME}). In particular, the RME amplitude of $\approx100\,\mathrm{m\,s}^{-1}$ is quite substantial. In comparison, the RME amplitude of HD209458\,b is $\approx40\,\mathrm{m\,s}^{-1}$ \citep{2000AaA...359L..13Q}, and recent RME detections probe down to $1\,\mathrm{m\,s}^{-1}$ \citep{2010ApJ...723L.223W}.

\section{Discussion and Conclusion}
\label{subsec:conclusion}

We present two new methods to determine an exomoon's sense of orbital motion. One method, which we refer to as the transit timing dichotomy, is based on a light travelling effect that occurs in subsequent mutual planet-moon eclipses during stellar transits. For the ten largest moons in the solar system, TTDs range between 2 and 12\,s. If the planet-moon orbital period can be determined independently (e.g. via TTV and TDV measurements) and with an accuracy of $\lesssim1\,\mathrm{hr}$, or if a very close-in moon shows at least three mutual events during one stellar transit, then TTD can uniquely determine the sequence of planet-moon eclipses. To resolve the TTD effect, photometric accuracies of $10^{-4}$ need to be obtained along with mid-event precisions $\lesssim6$\,s, which should be possible with the E-ELT. If an exomoon is self-luminous, e.g. due to extreme tidal heating \citep{2013ApJ...769...98P}, and shows regular planet-moon transits and eclipses every few days, its TTD might be detectable without the need of stellar transits. Eccentricities as small as 0.001 can trigger tidal surface heating rates on large moons that could be detectable with the James Webb Space Telescope \citep{2013ApJ...769...98P,2014arXiv1408.6164H}. But TTD measurements do not require such an extreme scenario, which does not support the conclusion of \citet{2014ApJ...791L..26L} that mutual events can only be used to determine an exomoon's SOM if the moon's own brightness can be measured. In general, TTDs can be used to verify the prograde or retrograde motion of a moon with respect to the circumstellar motion. 

The second method is based on measurements of the IR spectrum emitted by a young, luminous giant planet. We present simulations of the Rossiter-McLaughlin effect imposed by a transiting moon on a planetary spectrum. The largest moons that can possibly form in the circumplanetary accretion disks, halfway between Ganymede and Earth in terms of radii, can cause an RME that will be comfortably detectable with an IR spectrograph like CRIRES mounted to an ELT. A larger throughput or larger spectral coverage than the current non-cross-dispersed CRIRES will make it possible to determine the SOM of smaller moons, maybe similar to the ones we have in the solar system. A moon-induced planetary RME can determine the moon's orbital motion with respect to the planetary rotation.

Combined observations of an exomoon's TTD, its planetary RME as well as the moon's stellar RME \citep{2010MNRAS.406.2038S,2012ApJ...758..111Z} and the inclination between the moon's circumplanetary orbit and the planet's circumstellar orbit \citep{2011MNRAS.416..689K} can potentially characterize an exomoon orbit in full detail.

We conclude by emphasizing that a moon's transit probability in front of a giant planet is about an order of magnitude higher than that of a planet around a star. A typical terrestrial planet at 0.5\,AU from a Sun-like star has a transit probability of $R_\odot/0.5\,\mathrm{AU}\approx1\,\%$. For comparison, the Galilean moons orbit Jupiter at distances of about 6.1, 9.7, 15.5, and $27.2\,R_\mathrm{Jup}$, implying transit probabilities of up to about 16\%. With orbital periods of a few days, moon transits occur also much more frequently than for a common Kepler planet. Permanent, highly-accurate IR photometric monitoring of a few dozen directly imaged giant exoplanets thus has a high probability of finding an extrasolar moon.

\acknowledgments

The report of an anonymous referee was very helpful in clarifying several passages in this letter. We thank Tim-Oliver Husser for providing us with the PHOENIX models. Ren\'e Heller is supported by the Origins Institute at McMaster University and by the Canadian Astrobiology Program, a Collaborative Research and Training Experience Program funded by the Natural Sciences and Engineering Research Council of Canada (NSERC). Funding for the Stellar Astrophysics Centre is provided by The Danish National Research Foundation (Grant agreement no.: DNRF106).

%BIBLIOGRAPHY
%\bibliography{ms}
%\bibliographystyle{apj2}

\end{document}